\documentclass[11pt]{article}

\usepackage[numbers]{natbib}
\usepackage{graphicx}
\usepackage[labelfont=bf]{caption}

\usepackage{breakurl}
\usepackage{hyperref}
\hypersetup{colorlinks=true, citecolor=blue, linkcolor=blue, urlcolor=blue}

\usepackage[symbol]{footmisc}

\renewcommand{\section}{\subsection}

\begin{document}

\centerline{\bf\Large Nonprofit Adopt a Star: Lessons from 15 years of Crowdfunding}

\vspace*{11pt}
\centerline{Travis S. Metcalfe\footnote[1]{+1\,720-310-5180; 
\href{mailto:travis@wdrc.org}{travis@wdrc.org}} (White Dwarf Research Corporation)}

\begin{quote}
{\sl Summary:} In the past 15 years, the number of known planets outside of our solar 
system has grown from about 200 to more than 5000. During that time, we have conducted 
one of the longest crowdfunding campaigns in history, a nonprofit adopt a star program 
that supports astronomy research. The program includes the targets of NASA space 
telescopes that are searching for planets around other stars, and it uses the proceeds to 
help determine the properties of those stars and their planetary systems. I summarize how 
this innovative program has evolved over the years and engaged the public worldwide to 
support an international team of astronomers.
\end{quote}

\section*{Background}

In January 2008, we started a crowdfunding program known as the {\sl Pale Blue Dot 
Project} to support an international team of astronomers who were preparing for the 
launch of the Kepler space telescope the following year. The program offered an ``adopt a 
star'' service featuring the targets of the Kepler mission, with all of the proceeds 
supporting the work of the team to characterize the stars and their planetary systems 
\citep{1}. Donors received a personalized ``Certificate of Adoption'' by email, and their 
selected target was updated in our online database---ensuring that each star could only 
be adopted once. The database showed an image of the star in Google Sky, along with the 
constellation name and coordinates, a link to a star chart, and a link to additional 
information from the SIMBAD astronomical database \citep{2}. We have previously described 
our experiences during the first seven years of the program \citep{3}, so here we focus 
on developments after 2014.

\section*{Evolution of the Program}

To align the name of our program with the service that it provided, in 2014 we rebranded 
as {\sl Adopt a Star} (\href{https://adoptastar.org}{adoptastar.org}). This decision was 
motivated in part by a new generation of supporters who were not familiar with astronomer 
Carl Sagan, who coined the phrase ``pale blue dot'' to describe the Earth as viewed from 
a distance. The original name alluded to our search for pale blue dots around other 
stars, but the reference was obscure to many people. The ``adopt a star'' concept had 
previously been introduced \citep{4}, but it had not been actively publicized. We began 
promoting our adopt a star program in 2008 using keyword-based advertising sponsored by 
an in-kind grant from a leading search engine company. The phrase has now been embraced 
by deceptive ``name a star'' companies, to the extent that we recently rebranded our 
program {\sl Nonprofit Adopt a Star} to distinguish it from the for-profit competitors.

\begin{figure}[t]
\centerline{\includegraphics[angle=0,width=\textwidth,trim=0 150 0 0, clip]{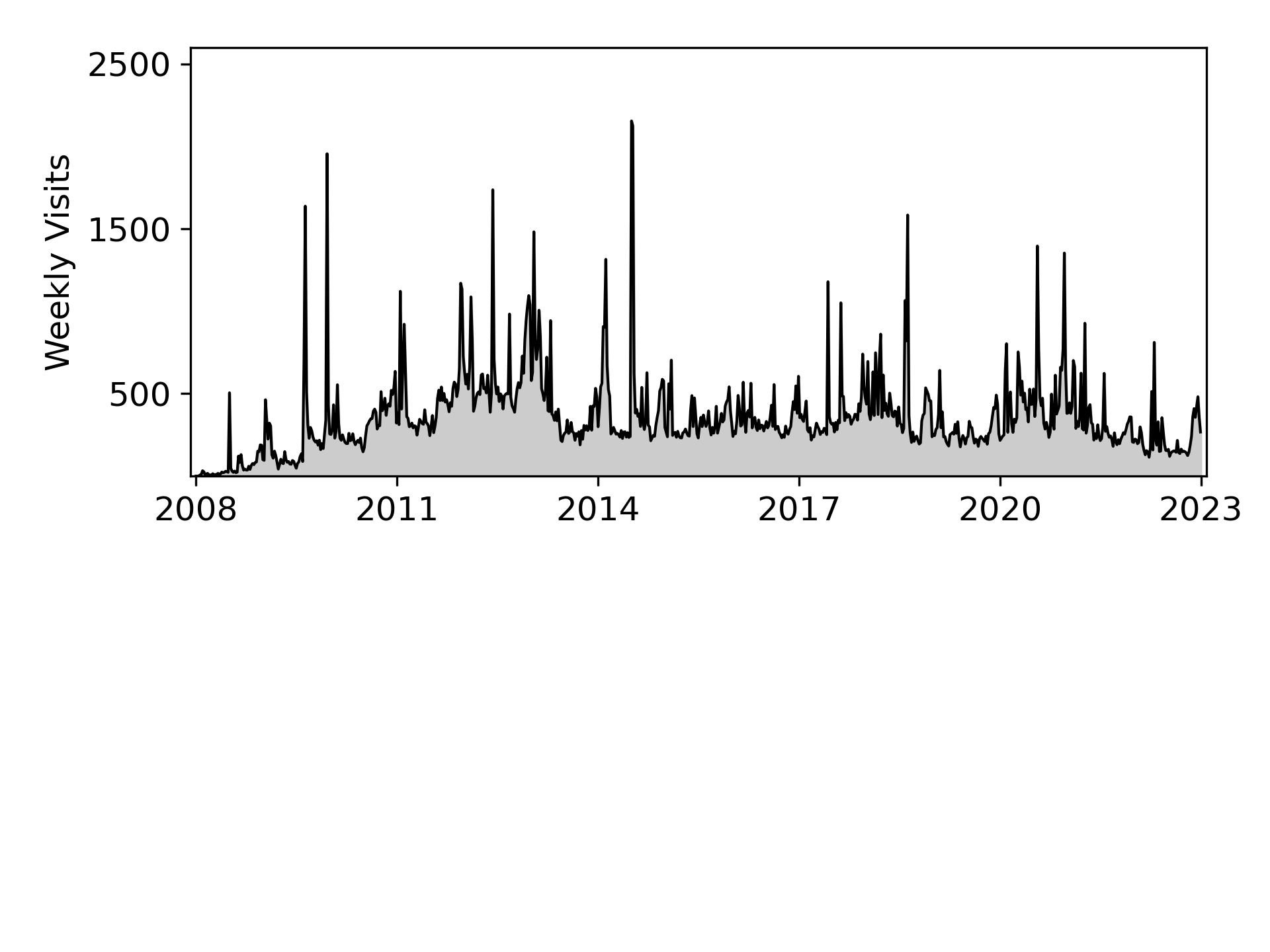}}
\caption{\sl Weekly visits during the 15 years of our crowdfunding campaign. Although 
visits stayed relatively constant after 2014, program revenue quadrupled as a higher 
fraction of visitors adopted stars and made larger donations for value-added 
targets.\label{fig1}}
\end{figure}

Our original adopt a star program included only the targets of the Kepler space 
telescope. The Kepler mission came to an abrupt end in 2013 after the spacecraft lost the 
second of four reaction wheels that provided pointing stability. However, some clever 
engineers realized that the two remaining wheels could still allow the telescope to 
observe stars in the ecliptic plane for up to a few months at a time \citep{5}. After 
demonstrating that it could work, the repurposed Kepler mission (known as K2) began 
operating in this mode in 2014, and went on to observe a series of 19 fields near the 
ecliptic plane over the next four years. Several hundred thousand targets were selected 
from a newly-created Ecliptic Plane Input Catalog \citep{6}, including many stars that 
are visible to the unaided eye. Our adopt a star program initially offered all targets 
for \$10, but we began providing value-added targets for larger donations in 2012: \$15 
for double stars, \$25 for stars with suspected planets, and \$50 for confirmed planetary 
systems. The K2 mission gave us the first opportunity to add bright stars (visible 
without a telescope) for a \$100 donation. The launch of the Transiting Exoplanet Survey 
Satellite (TESS) in 2018 extended this opportunity to bright stars in almost every 
constellation in the sky \citep{7}.

Although only 10 percent of our donors chose to adopt a bright star, the option 
eventually yielded half of the total revenue from the program. While we had attracted 
about \$100,000 in the first seven years of the {\sl Pale Blue Dot Project}, we raised an 
additional \$450,000 over the next eight years under the {\sl Adopt a Star} brand. As 
illustrated in Figure~\ref{fig1}, weekly visits to our website stayed relatively constant 
during this time, so the increase in revenue probably resulted from converting a higher 
fraction of our visitors into donors, and from convincing supporters to donate more for 
value-added targets. We believe that improvements to our website inspired more visitors 
to adopt a star, primarily a streamlining of the donation process (allowing the donor to 
select the brightest available star in any category by default) and enhancements of the 
product that we provided (better certificate design and additional features in the Google 
Sky interface through our database). We recently launched a refreshed website 
(\href{https://adoptastar.org}{adoptastar.org}) that amplifies these advantages, and 
couples the website and database into a more unified user experience.

\begin{figure}[t]
\centerline{\includegraphics[angle=0,height=4.25in]{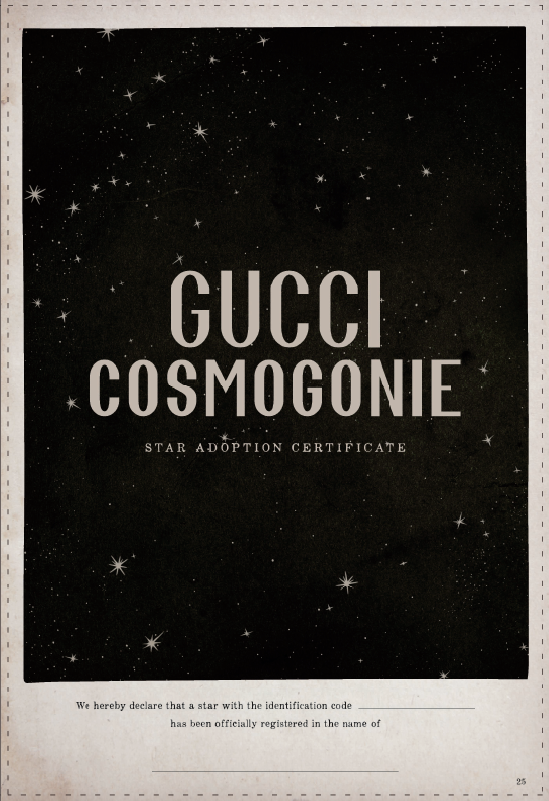}\hspace*{12pt}\includegraphics[angle=0,height=4.25in]{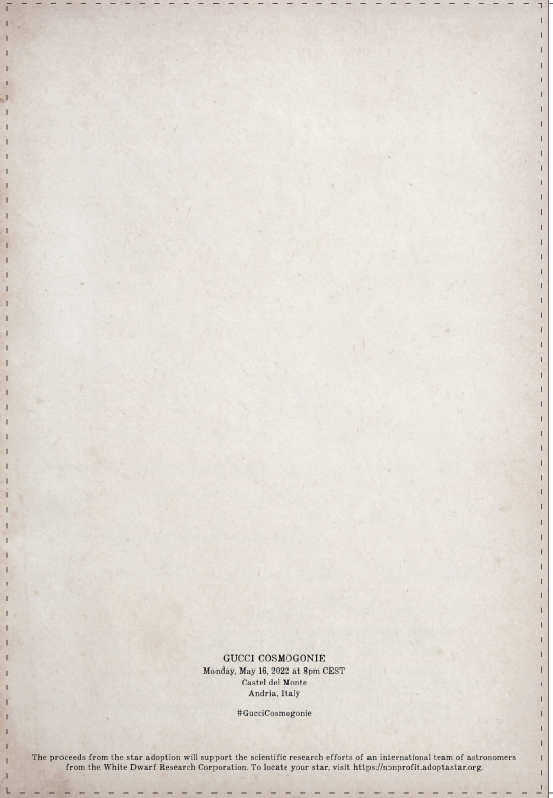}}
\caption{\sl Custom star adoption certificate designed for the Gucci Cosmogonie fashion 
show in May 2022. A bright star in a zodiac constellation was adopted for each of several 
hundred guests, yielding the largest single donation in the history of our crowdfunding 
campaign.\label{fig2}}
\end{figure}

\section*{New Successes \& Challenges}

We have previously described \citep{3} how there is an interesting story behind almost 
every peak in Figure~\ref{fig1}. In addition to the recurring features that correspond to 
popular gift-giving holidays (Christmas and Valentine’s Day), there are episodic spikes 
that coincide with traditional media coverage or social media exposure. A few of these 
successes triggered some of our early challenges, including institutional pressure from 
NASA, threats of legal action by the estate of Carl Sagan, and an insult to Russian 
president Vladimir Putin by Ukrainian astronomers \citep{8}. More recent spikes have been 
largely positive, tapping into worldwide enthusiasm for our program. A large spike in 
August 2018 corresponds to a surge in star adoptions for Qixi festival (Chinese 
Valentine’s Day). A peak in mid-2020 resulted from a birthday star adoption for Filipino 
social media influencer Bretman ``Da Baddest'' Rock, who featured the gift on Instagram. 
A surge of visitors in early 2021 arrived on the fifteenth anniversary of Twitter, when a 
marketing company released a website featuring the top tweets of 140 people in the tech 
Twitterverse, with a blue star adoption in the Pleiades cluster for each person 
(including Elon Musk).

The greatest success of our {\sl Nonprofit Adopt a Star} program came in the spring of 
2022, when we were contacted by the press office of Gucci in Italy. They wanted to 
discuss a collaboration opportunity, so they set up a video call for later in the day. 
Ahead of the call they asked us to sign a non-disclosure agreement, citing the discussion 
of highly confidential information. It turned out that Gucci was organizing a 
space-themed fashion show at a castle in Italy \citep{9}, and they wanted to adopt a 
bright star in a zodiac constellation for each of their several hundred guests. We worked 
with the team to coordinate the star adoptions, ensuring that none of the details 
appeared in our database until the day of the fashion show. Gucci designed the custom 
star adoption certificate shown in Figure~\ref{fig2}, which was included in a gift bag 
for each of their guests. The resulting payment was the largest single donation in the 
history of our crowdfunding program, representing nearly half of the support that we had 
received in the previous year.

All proceeds from our {\sl Nonprofit Adopt a Star} program support the research efforts 
of the Kepler/TESS Asteroseismic Science Consortium (\href{https://tasoc.dk}{tasoc.dk}), 
a large collaboration led by scientists in Denmark. The leadership of this organization 
has generally tolerated our crowdfunding campaign, and one of the greatest challenges has 
been to get the broader membership to invest in the concept. The collaboration as a whole 
has gladly accepted our co-sponsorship of annual science conferences, which has given 
many students and early career researchers the opportunity to attend and present their 
work without paying a registration fee. Some individual members in developing countries 
have also accepted support for the publication charges on their scientific papers, 
helping them publish in top-tier journals. But efforts to engage the collaboration in 
direct promotion of the campaign for their own benefit have largely failed. This may 
reflect a difference in the culture of philanthropy in European countries, where 
investment in scientific research is seen as the responsibility of governments rather 
than individuals, and where there are relatively few incentives for charitable giving.

\section*{Future Outlook}

Considering the history of our well-established brand, there is enormous potential for 
current and future planet-search missions to benefit from our crowdfunding program. The 
TESS mission is approved for continued operations through September 2025, and NASA has 
already invited the mission to propose an additional extension through 2028. The European 
PLATO mission is currently scheduled for launch in late 2026, and the baseline observing 
plan includes two years of monitoring for two different fields that each cover an area 20 
times larger than the original Kepler field 
(\href{https://platomission.com}{platomission.com}). This mission will certainly rely on 
international teams to help analyze the observations, but there is presently no NASA 
program to support the participation of U.S.\ scientists. With cooperation from the PLATO 
mission, citizens worldwide will be able to support the next wave of discoveries through 
{\sl Nonprofit Adopt a Star}.\\

\noindent{\sl Acknowledgments: White Dwarf Research Corporation receives in-kind support 
from Google, Stripe, and Dreamhost. We would like to thank Sjors Provoost and Robert 
Piller for contributions that substantially improved our crowdfunding program.}

\end{document}